\documentclass[12pt]{article}
\usepackage{amsmath, amsfonts,euscript,bbm}
\usepackage{epsfig, cite}
\usepackage{color}
\usepackage{ifthen}
\usepackage{graphicx}
\usepackage{setspace} 
\newboolean{Notes}\setboolean{Notes}{true}
\newcommand{\notes}[1]{\ifthenelse{\boolean{Notes}}{\textcolor{black}{#1}}{}}

\def\be{\begin{equation}}
\def\ee{\end{equation}}
\def\bea{\begin{eqnarray}}
\def\eea{\end{eqnarray}}

\newcommand{\skyp}[1]{}

\begin{document}

\bigskip
\hskip 5in\vbox{\baselineskip12pt \hbox{FNAL-PUB-08-215-A-T} }
\bigskip\bigskip\bigskip\bigskip\bigskip

\centerline{\Large Gravitational Wave Bursts from Cosmic Superstring Reconnections}
\bigskip
\bigskip
\bigskip
\centerline{\bf Mark G. Jackson$^{1,2}$ and Xavier Siemens$^3$}
\medskip
\centerline{$^1$Particle Astrophysics Center and Theory Group}
\centerline{Fermi National Accelerator Laboratory}
\centerline{Batavia, Illinois 60510, USA}
\centerline{$^2$Lorentz Institute for Theoretical Physics}
\centerline{Leiden 2333CA, The Netherlands}
\centerline{\it markj@lorentz.leidenuniv.nl}
\medskip
\centerline{$^3$Center for Gravitation and Cosmology}
\centerline{Department of Physics}
\centerline{University of Wisconsin - Milwaukee}
\centerline{P.O. Box 413}
\centerline{Wisconsin, 53201, USA}
\centerline{\it siemens@gravity.phys.uwm.edu}
\bigskip
\bigskip
\begin{abstract}
  We compute the gravitational waveform produced by cosmic superstring
  reconnections. This is done by first constructing the superstring
  reconnection trajectory, which closely resembles that of classical,
  instantaneous reconnection but with the singularities smoothed out
  due to the string path integral.  We then evaluate the graviton
  vertex operator in this background to obtain the burst amplitude.
  The result is compared to the detection threshold for current and
  future gravitational wave detectors, finding that neither bursts nor
  the stochastic background would be detectable by Advanced LIGO.
  This disappointing but anticipated conclusion holds even for the most optimistic
  values of the reconnection probability and loop sizes.
\end{abstract}

\newpage            
\baselineskip=18pt
\section{Introduction}

One of the most exciting products of the recent synthesis of
superstring theory and cosmology has been that of cosmic superstrings
\cite{Polchinski:2004ia}.  Like their classical cousins
\cite{Kibble:1976sj}, cosmic superstrings were briefly considered then
discarded \cite{Witten:1985fp}, but for theoretical reasons rather
than observational ones.  Subsequent (and largely non-perturbative)
analysis \cite{nonpert} has shown that cosmic superstrings have been
found to naturally arise in models of brane inflation \cite{cstrings},
a string theory realization of inflationary cosmology theory.

Cosmic strings and superstrings can produce a variety of astrophysical
signatures including gravitational
waves~\cite{dandv,Siemens:2006vk,Hogan:2006we,Siemens:2006yp,DePies:2007bm}, ultra high energy
cosmic rays~\cite{berez}, and gamma ray bursts~\cite{bands}. Very
recent work has revealed a number of exciting new possibilities such
as radio bursts from strings~\cite{Vachaspati:2008su}, effects on the
cosmic 21~cm power spectrum~\cite{Khatri:2008zw},
magnetogenesis~\cite{Battefeld:2007qn}, effects on the CMB at small
angular scales~\cite{Fraisse:2007nu,Pogosian:2008am}, CMB polarization \cite{Baumann:2008aj}, microlensing
from strings~\cite{Chernoff:2007pd,Kuijken:2007ma}, strong
lensing~\cite{Gasparini:2007jj,Christiansen:2008vi}, and weak
lensing~\cite{Dyda:2007su}.

Prior work on gravitational radiation has focused on two processes:
cusps (whereby a segment of the string momentarily moves at the speed
of light) and kinks (formed after two cosmic strings collide and
reconnect)~\cite{dandv,Siemens:2006vk,Hogan:2006we,Siemens:2006yp,DePies:2007bm}.
Here we study a third source: the radiation emitted from the
reconnection process itself.  This is possible for cosmic superstrings
because string theory allows us to explicitly construct the
reconnection process and compute detailed interaction properties.  We
will do this for the bosonic string but the presence of fermions in
the superstring should not change the conclusions.

In the first half of this article we compute the gravitational
waveform resulting from fundamental cosmic superstring reconnection.
In the second half we consider the likelihood of detection of this
signal with current and future experiments.

\section{Cosmic Superstring Reconnection}
\subsection{Basic Process}
\begin{figure}
\begin{center}
\includegraphics[width=6in]{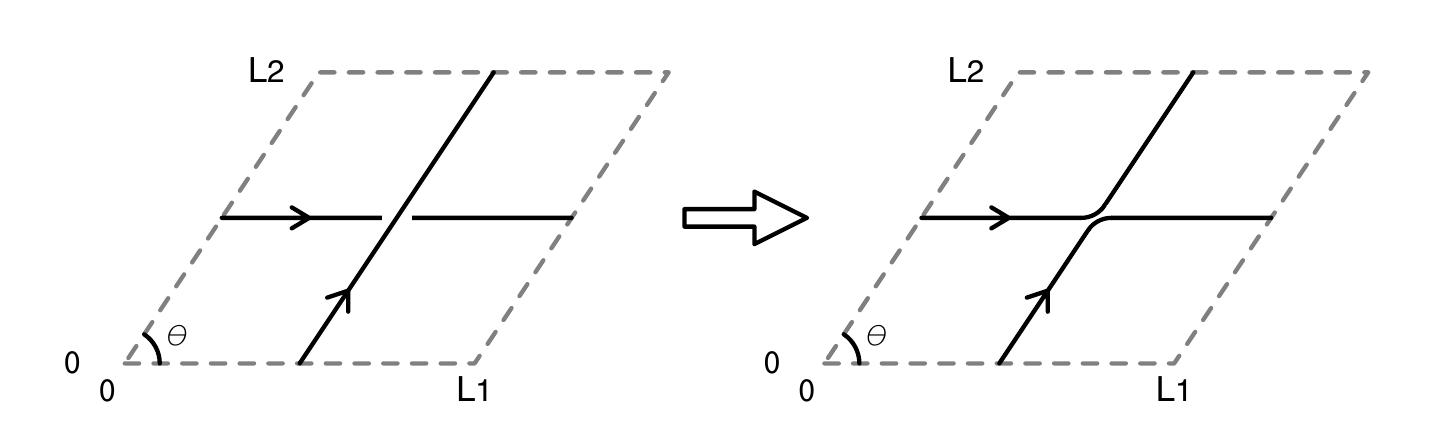}
\caption{We model cosmic superstrings as straight wound modes on a large torus, which will then interact to form a kinked configuration.} \ \\
\end{center}
\end{figure}
The theory of cosmic superstring reconnection was developed in \cite{Polchinski:1988cn}  \cite{Jackson:2004zg} \cite{Jackson:2007hn}.  Consider two long, straight wound bosonic strings on a 2D torus of size $L$ and skew angle $\theta$ as illustrated in Figure 1.  In terms of the string tension $(2 \pi \alpha')^{-1}$ the momenta are taken to be 
\begin{equation}
\label{pl}
p_1 = \left[ \left( \frac{L}{2 \pi \alpha'} \right)^2 - \frac{4}{\alpha'} \right]^{1/2} (1,0,0,0, {\bf 0} ), \hspace{0.5in} L_1 = L (0,1,0,0, {\bf 0}),
\end{equation}
\[ p_2 = \left[ \left( \frac{L}{2 \pi \alpha'} \right)^2 - \frac{4}{\alpha'} \right]^{1/2} [1-v^2]^{-1/2} (1,0,0,v, {\bf 0} ), \hspace{0.5in} L_2 = L (0,\cos \theta,\sin \theta,0, {\bf 0}) . \]
These satisfy the tachyonic mass-shell conditions
\[ p^2_{iL} = p^2_{iR} = \frac{4}{\alpha'}, \hspace{0.5in} p_{iL/R} = p_i \pm \frac{L_i}{2 \pi \alpha'} . \]
The relevant vertex operators are: ($i=1,2$)
\begin{equation}
\label{simplevo}
 V_T(z, {\bar z}; p_i) = \frac{\kappa}{2 \pi \sqrt{V}} :e^{ip_{iL} X_L(z) + ip_{iR} X_R ({\bar z})}:
 \end{equation}
where the volume $V = V_\perp L^2 \sin \theta $ is the product of the the transverse volume and the 2D torus (methods to calculate $V_\perp$ can be found in \cite{Jackson:2006rb}).  Here $X_L(z), X_R({\bar z})$ refer to the (anti)holomorphic components of $X(z, {\bar z})$, so that $X(z,{\bar z}) = X_L(z) + X_R({\bar z})$.  

These will scatter into some kinked configuration and the amplitude for all such processes can be summed.  In the large-winding limit the probability of reconnection is found to be
\begin{eqnarray}
\nonumber
P &=& \frac{1}{4 E_1 E_2 v} \int \frac{d {\bf p}_{f}}{2 \pi} \frac{1}{2E_f} \sum_f |\mathcal M_f|^2 (2 \pi )^2 \delta^2(p_1 + p_2 - p_f) \\
\label{p0}
&=& g_s^2 \frac{V_{\rm min} }{V_\perp}  \frac{ (1- \cos \theta \sqrt{1-v^2})^2}{8 \sin \theta v \sqrt{1-v^2}},  \hspace{0.5in} V_{\rm min} = (4 \pi^2 \alpha')^3 . 
\end{eqnarray}

It is also likely that there will be radiation emitted during this reconnection.  This can be included by using the reconnection process as a classical background trajectory $X_{\rm cl}(z, {\bar z})$ upon which the radiation vertex operator must be integrated over \cite{Jackson:2007hn}.  Then the probability to emit a radiated state of definite momentum $k$ will be the probability of reconnection times a factor depending on the radiated particle under consideration,
 \begin{equation}
 \label{prad}
  P_{\rm rad} (k) = P_0 \left( \frac{ \alpha' g^2_s V_{\rm min}}{(4 \pi)^4 V_\perp } \right) \left| \int d^2 z \ V_{\rm rad} [k;X_{\rm cl}(z, {\bar z})] \right|^2 .
  \end{equation}
One might also consider the possibility that the strings will pass through each other without reconnecting but still emit radiation in the process.  This is a valid possibility but one in which the amplitude is suppressed by a factor of $g_s$ due to three final strings instead of two (recall that a reconnected kinked string is considered a single string, whereas this process will produce two unconnected strings with accompanying coupling factor).  So to leading order this process can be neglected.   
\subsection{Computing the Background Trajectory}
The reconnection trajectory was computed in \cite{Jackson:2007hn} by considering the vertex operator of the kinked string.  This is done by examining the operator product expansion of the straight strings,
\begin{eqnarray*}
\label{kinkvertex}
: e^{ip_{1L} X_L(z)}::e^{ip_{2L} X_L(0)}: &=& z^{ {\alpha' \over 2} p_{1L} \cdot p_{2L} } : e^{ip_{1L}X_L(z)+i p_{2L} X_L(0)}: \\
&=& z^{ {\alpha' \over 2} p_{1L} \cdot p_{2L} }: \left( 1 + i z p_{1L} \cdot \partial X_L(0) + \ldots \right) e^{i(p_{1L}+p_{2L})X_L(0)}: .
\end{eqnarray*}
The Taylor expansion of the exponential shows the vertex operators of the infinite tower of the produced states, which will appear kinked due to their large oscillator excitation number $N$:
\begin{eqnarray*}
N-1 &=& - \frac{\alpha'}{4} (p_{1L}+p_{2L})^2 \\
&=& -2 - \frac{\alpha'}{2} p_{1L} \cdot p_{2L} \\
&\sim& L^2 / \alpha' .
\end{eqnarray*}
Since $p_{1R} \cdot p_{2R} = p_{1L} \cdot p_{2L}$ the result will be identical for the right-moving oscillators and so ${\tilde N}=N$.  Now consider the state corresponding to the sum of these vertex operators (\ref{kinkvertex}):
\begin{eqnarray*}
| {\rm kinks} \rangle &=&  \frac{1}{(2 \pi i)^2}  \oint_0 d \epsilon d {\bar \epsilon} \ |\epsilon|^{-2(N+1)} V_T(\epsilon, {\bar \epsilon}; p_1) V_T(0; p_2) \\
&=& \frac{1}{(2 \pi i)^2}  \oint_0 d \epsilon d {\bar \epsilon} \ |\epsilon|^{-2(N+1)} e^{ \sqrt{\alpha' \over 2} \sum_{n \geq 1} p_{1L} \cdot \alpha_{-n} \epsilon^n / n+ p_{1R} \cdot {\tilde \alpha}_{-n} {\bar \epsilon}^n / n} |p_{1}+p_{2} \rangle. 
\end{eqnarray*}
The expectation value of the position operator in the reconnection amplitude is then easily evaluated:
\begin{eqnarray*}
\nonumber
X_{\rm cl}(z, {\bar z}) &=& \frac{ \langle V_{\rm kinks}(\infty;p_1,p_2)  X(z, {\bar z}) V_T(1;p_1) V_T(0; p_2) \rangle}{\langle  V_{\rm kinks}(\infty;p_1,p_2) V_T(1;p_1) V_T(0; p_2) \rangle} \\
&=&  -i {\alpha' \over 2} p_{2L} \ln z -i {\alpha' \over 2} p_{1L} \left[ \ln (z-1) + \sum _{n=1}^N \left( \frac{1}{n} - \frac{1}{N+1} \right) z^n \right] \ + \ (L \rightarrow R, z \rightarrow {\bar z}). 
 \end{eqnarray*}
While this is a technically correct answer, it is unsatisfactory in that it appears to treat string 1 different from string 2 even though of course these are unphysical labels.  The reason is that the placement of vertex operators has treated them in a non-symmetric fashion, expanding $V_T(\epsilon; p_1)$ around the location of $ V_T(0; p_2)$.  To remedy this we simply redo the previous calculation but expand both vertex operators around the midpoint of their separation,
\begin{eqnarray}
\label{kinkstate}
| {\rm kinks} \rangle &=&  \frac{1}{(2 \pi i)^2}  \oint_0 d \epsilon d {\bar \epsilon} \ |\epsilon|^{-2(N+1)} V_T(\epsilon/2; p_1) V_T(-\epsilon/2; p_2) \\
\nonumber
&=& \frac{1}{(2 \pi i)^2}  \oint_0 d \epsilon d {\bar \epsilon} \ |\epsilon|^{-2(N+1)} e^{ \sqrt{\alpha' \over 2} \sum_{n \geq 1} p_{1L} \cdot \alpha_{-n} (\epsilon/2)^n / n+ p_{2L} \cdot \alpha_{-n} (-\epsilon/2)^n / n  + (L \rightarrow R) } |p_{1}+p_{2} \rangle. 
\end{eqnarray}
A good check that this is a physically equivalent solution (i.e. obtainable via a conformal transformation $z'(z)$) is that for the massless kink state ($N=1$) the polarization vector $p_{1L,R}$ has merely been replaced with $(p_{1L,R} - p_{2L,R})/2$, which is a difference proportional to the wavevector $p_{1L,R}+p_{2L,R}$ and thus pure gauge.  Now combining this new kink vertex operator with symmetric relocations $z=\pm \frac{1}{2}$ for the incoming straight-string vertex operators, the transformed trajectory is
\begin{eqnarray*}
X_{\rm cl}(z, {\bar z}) &=& \frac{ \langle V_{\rm kinks}(\infty;p_1,p_2)  X(z, {\bar z}) V_T(\frac{1}{2};p_1) V_T(-\frac{1}{2}; p_2) \rangle}{\langle V_{\rm kinks}(\infty;p_1,p_2) V_T(\frac{1}{2};p_1) V_T(-\frac{1}{2}; p_2) \rangle} \\
 &=&  - i {\alpha' \over 2} p_{1L} \left[ \ln \left( z-\frac{1}{2} \right) +  \sum _{n=1}^N \left( \frac{1}{n} - \frac{1}{N+1} \right) \left( \frac{z}{2} \right) ^n \right] \\
 &-&  i {\alpha' \over 2} p_{2L} \left[ \ln \left(z+ {1 \over 2} \right) + \sum _{n=1}^N \left( \frac{1}{n} - \frac{1}{N+1} \right) \left( - \frac{z}{2} \right) ^n \right] \ + \ (L \rightarrow R, z \rightarrow {\bar z}). 
 \end{eqnarray*}
As hoped, this is symmetric under $p_{1L} \leftrightarrow p_{2L}$ by exchanging $z \leftrightarrow -z$ (up to subtleties involving branch cuts).  This simplifies in the cosmic string limit ($N \rightarrow \infty$) by performing the summation,
\[ \sum _{n=1}^\infty \left( \frac{1}{n} - \frac{1}{N+1} \right) \left( \pm \frac{z}{2} \right) ^n \rightarrow - \ln \left( 1 \mp \frac{z}{2} \right) - \frac{1}{N+1} \frac{1}{1 \mp \frac{z}{2}}. \]
We will neglect the second term since this is a good approximation for most of the parameter space, and the trajectory simplifies tremendously:
\begin{equation}
\label{xcltemp}
X_{\rm cl}(z, {\bar z}) =  - i {\alpha' \over 2} p_{1L} \ln \left( \frac{2 z-1}{2 - z} \right)- i {\alpha' \over 2} p_{2L} \ln \left( \frac{2 z+1}{2 + z} \right) + (L \rightarrow R, z \rightarrow {\bar z}). 
\end{equation}
However, the trajectory (\ref{xcltemp}) as written is complex, due to the analytic continuation required by string theory; we now need to transform this back into a trajectory in real space.  This is easily done by recalling that vertex operators are inserted into the path integral as source terms via
\[ S = - \frac{1}{ 2 \pi \alpha'} \int d^2 z \ | \partial X|^2 - i p_L X_L(z_0) - i p_R X_R ({\bar z}_0)  . \]
This produces a trajectory
\begin{eqnarray*}
X(z, {\bar z}) &=& - i \frac{\alpha'}{2} p_L \ln (z-z_0) -  i \frac{\alpha'}{2} p_R \ln ( {\bar z}-{\bar z}_0) \\
&=& - i \frac{\alpha'}{2} p \ln |z-z_0|^2 - i \frac{L}{4 \pi} \ln \frac{z-z_0}{ {\bar z} - {\bar z}_0}.
\end{eqnarray*} 
We see the winding term is real, and that moving around the vertex operator produces $X \rightarrow X + L$ as desired.  The momenta term, however, is purely imaginary; this can be remedied by rotating $p \rightarrow ip$, allowing us to interpret the (logarithmic) distance away from the vertex operator as displacement in time.  Thus the trajectory for the reconnection process in real target space is
\begin{eqnarray}
\label{xcl}
X^\mu_{\rm cl}(z, {\bar z}) &=& {\alpha' \over 2} \left[ p_{1}^\mu \ln \left| \frac{2 z-1}{2 - z} \right|^2 + p_{2}^\mu \ln \left| \frac{2 z+1}{2 + z} \right|^2  \right]  \hspace{1in} \\
\nonumber
 &&  \hspace{1in} + \frac{1}{4 \pi i} \left[ L_{1}^\mu \ln \left( \frac{2 z-1}{2 {\bar z}-1} \cdot \frac{2 - {\bar z}}{2 - z} \right) + L_{2}^\mu  \ln  \left( \frac{2 z+1}{2 {\bar z}+1} \cdot \frac{2 + {\bar z}}{2 + z} \right) \right].
 \end{eqnarray}
This  reconnection process is shown in Figure 2, from the worldsheet and spacetime viewpoints.  The infinite past (containing the two incoming asymptotic states of simple wound strings) are located at $z = \pm \frac{1}{2}$, the infinite future (containing the single outgoing asymptotic state of a sharply kinked reconnected string) is located at $z = \pm 2$, and the entire unit circle $|z|=1$ maps to time $X^0=0$.  As expected, string theory has smoothed out what is classically a very violent and singular process. 
\begin{figure}
\begin{center}
\includegraphics[width=1.2in]{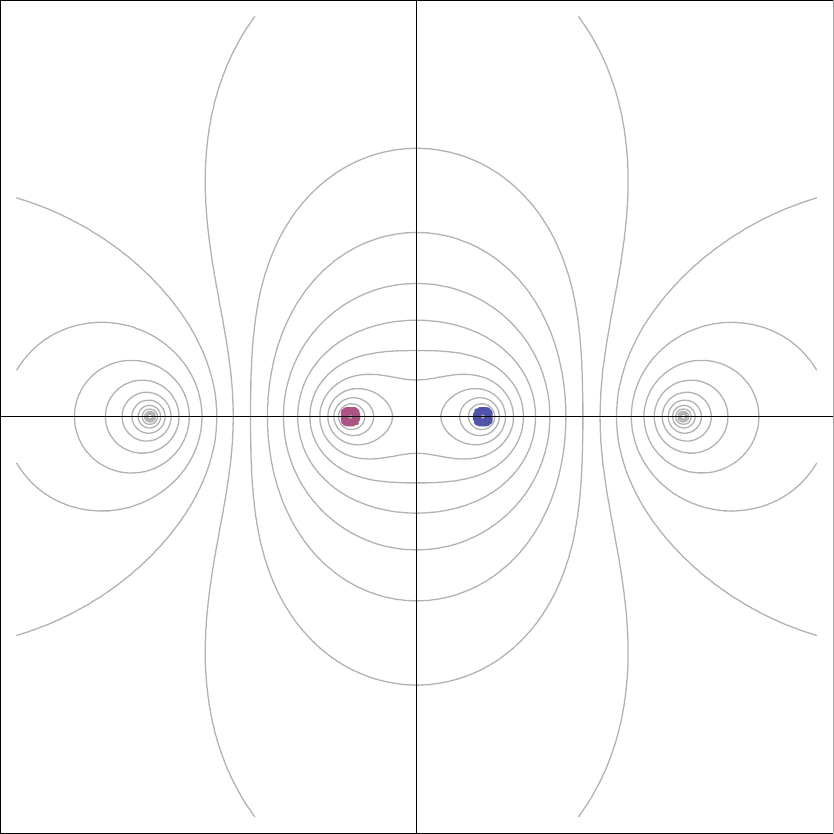}
\includegraphics[width=1.2in]{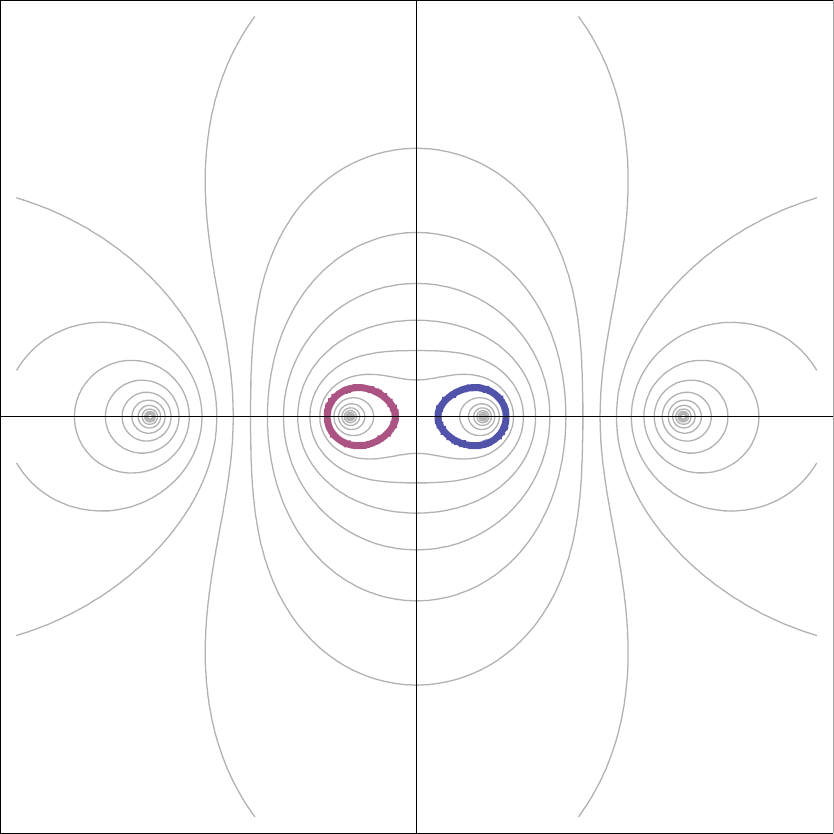}
\includegraphics[width=1.2in]{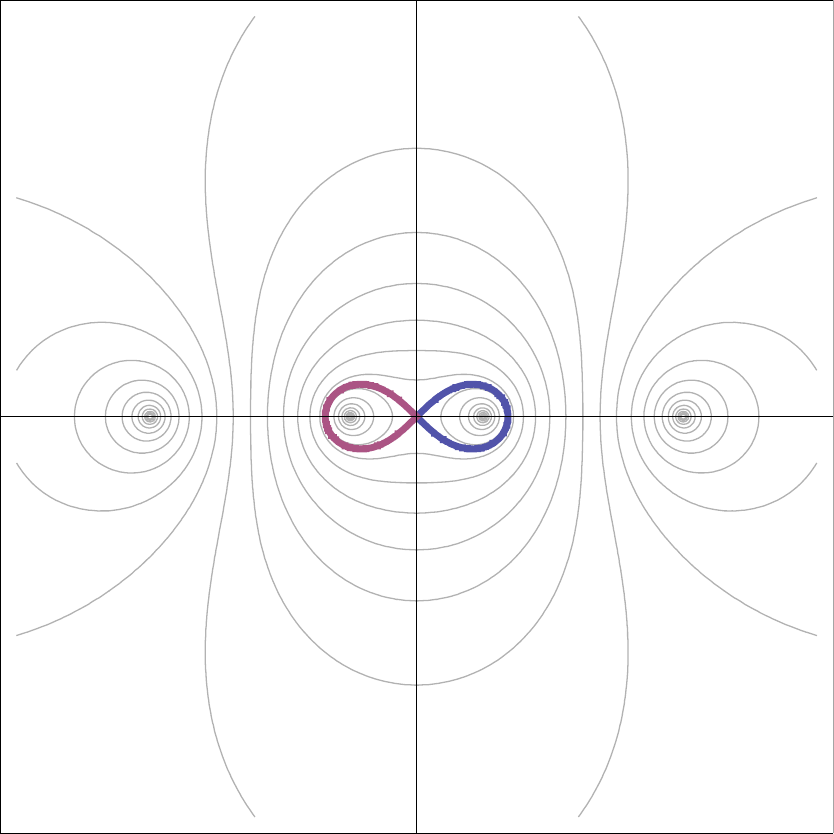} \\
\includegraphics[width=1.2in]{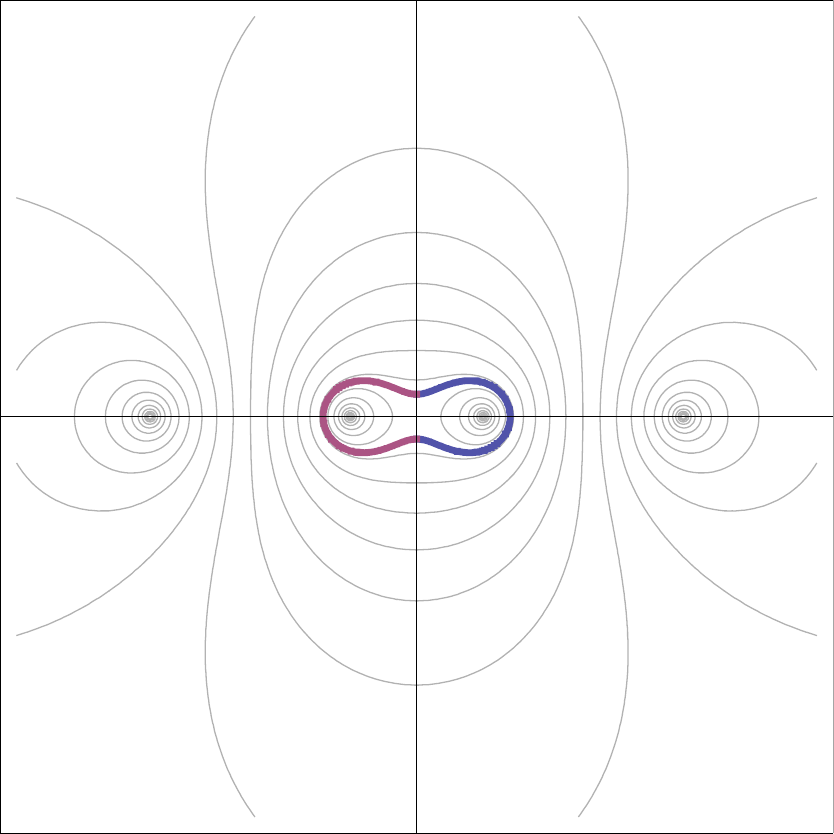}
\includegraphics[width=1.2in]{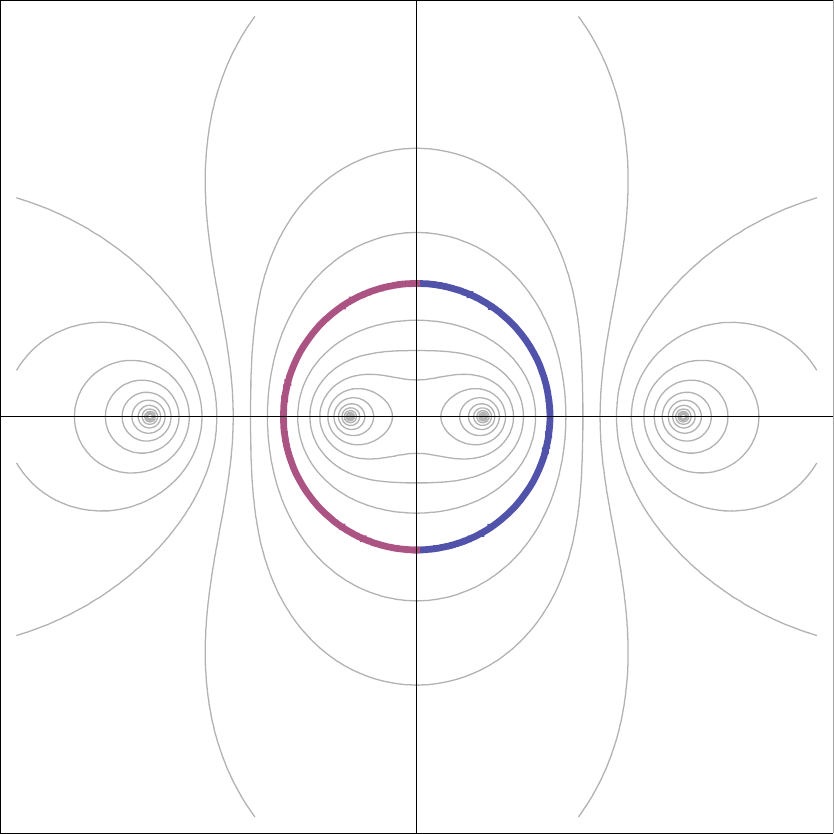}
\includegraphics[width=1.2in]{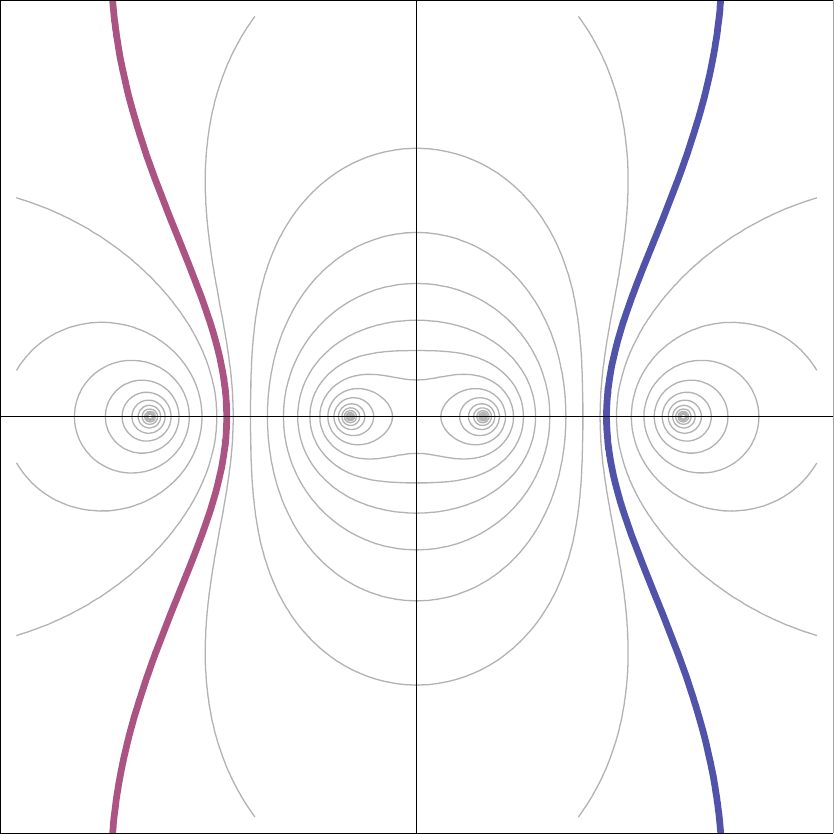} \\
\includegraphics[width=1.2in]{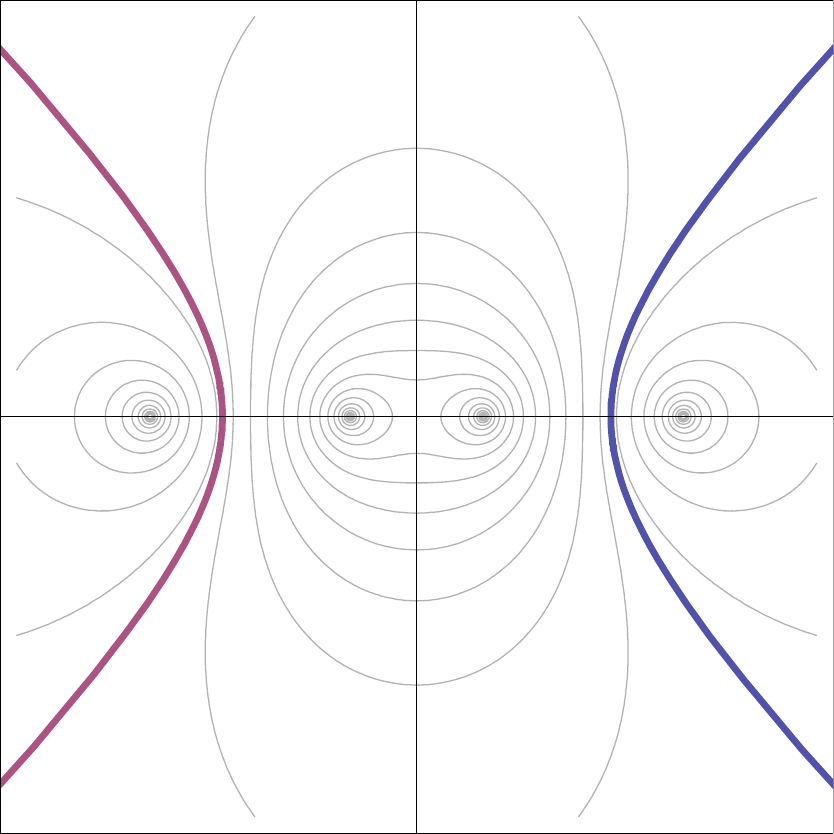}
\includegraphics[width=1.2in]{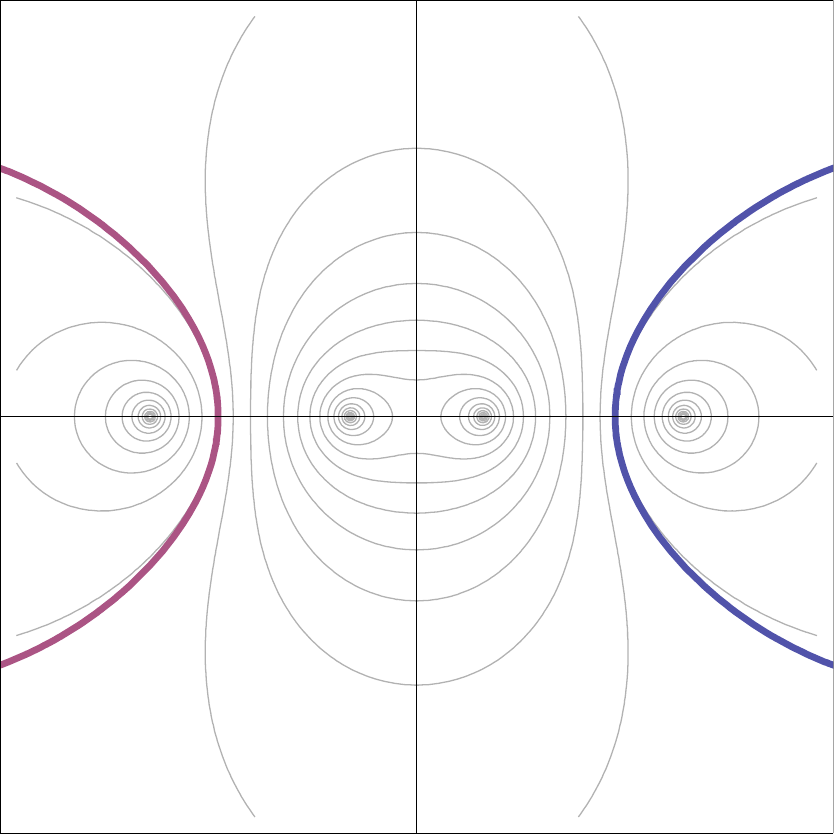}
\includegraphics[width=1.2in]{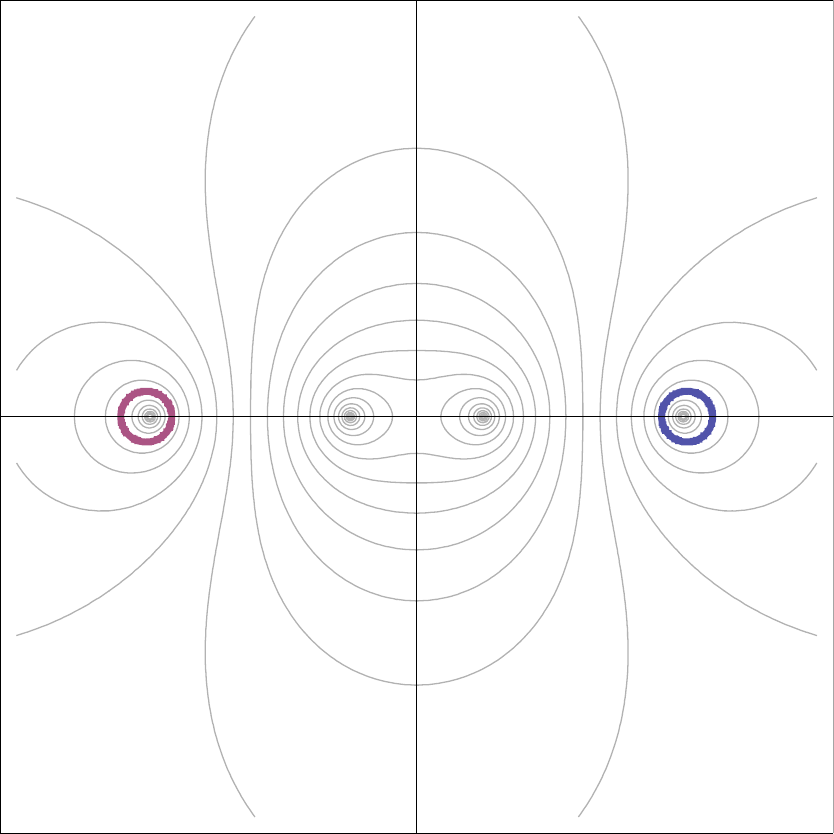} \\
\includegraphics[width=1.2in]{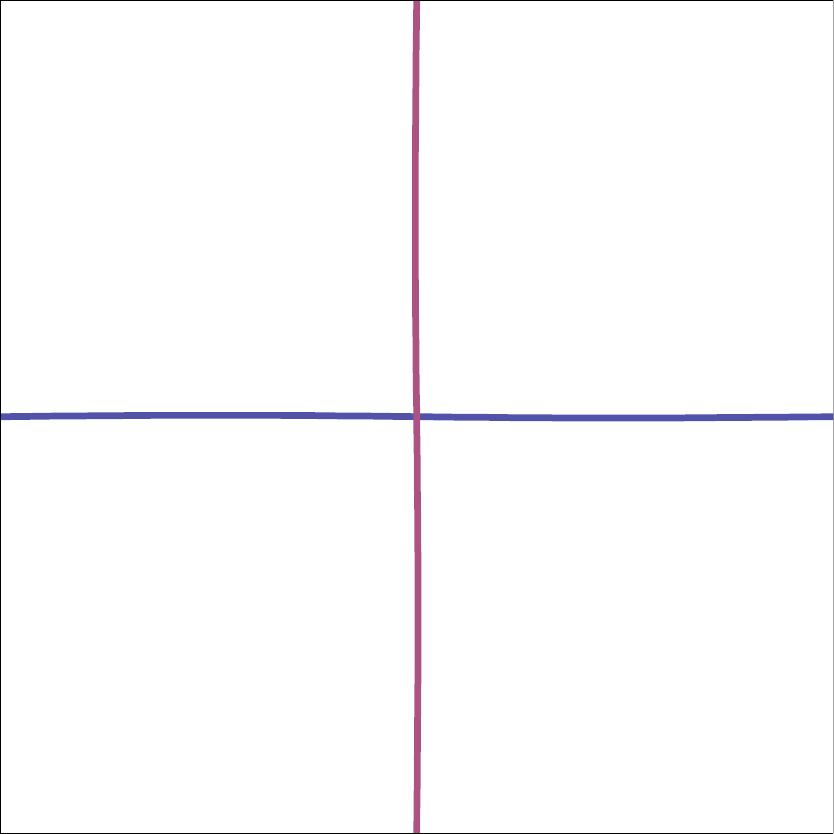}
\includegraphics[width=1.2in]{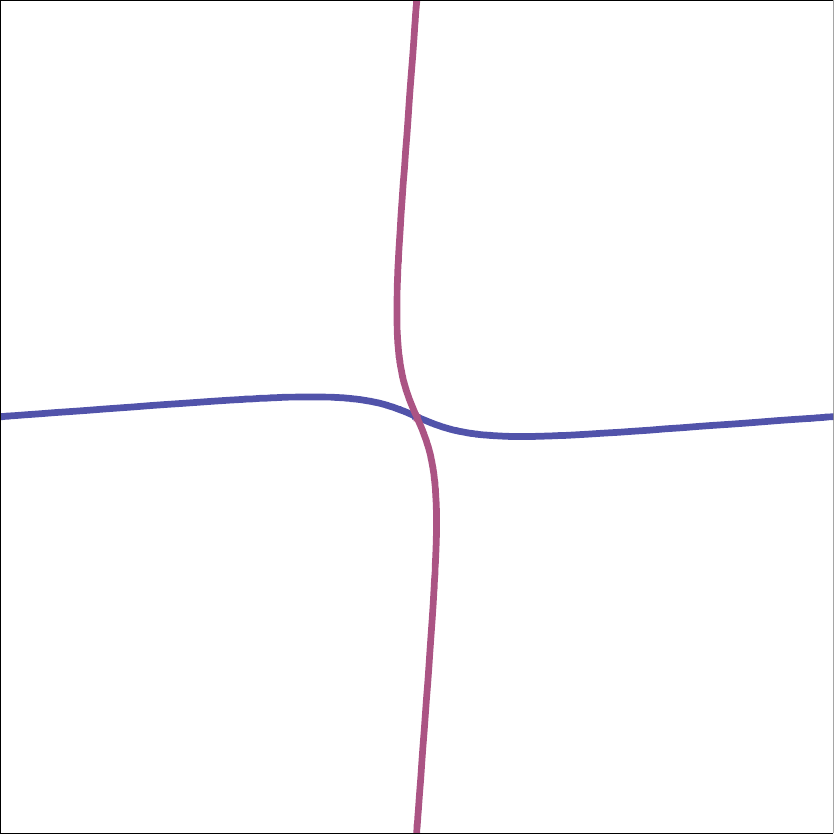}
\includegraphics[width=1.2in]{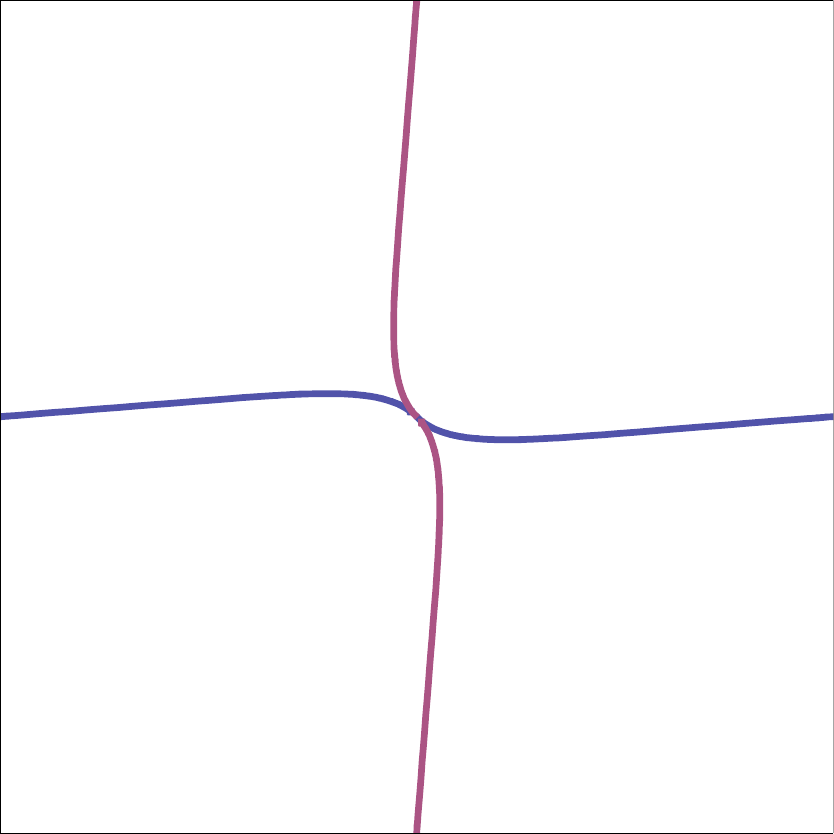} \\
\includegraphics[width=1.2in]{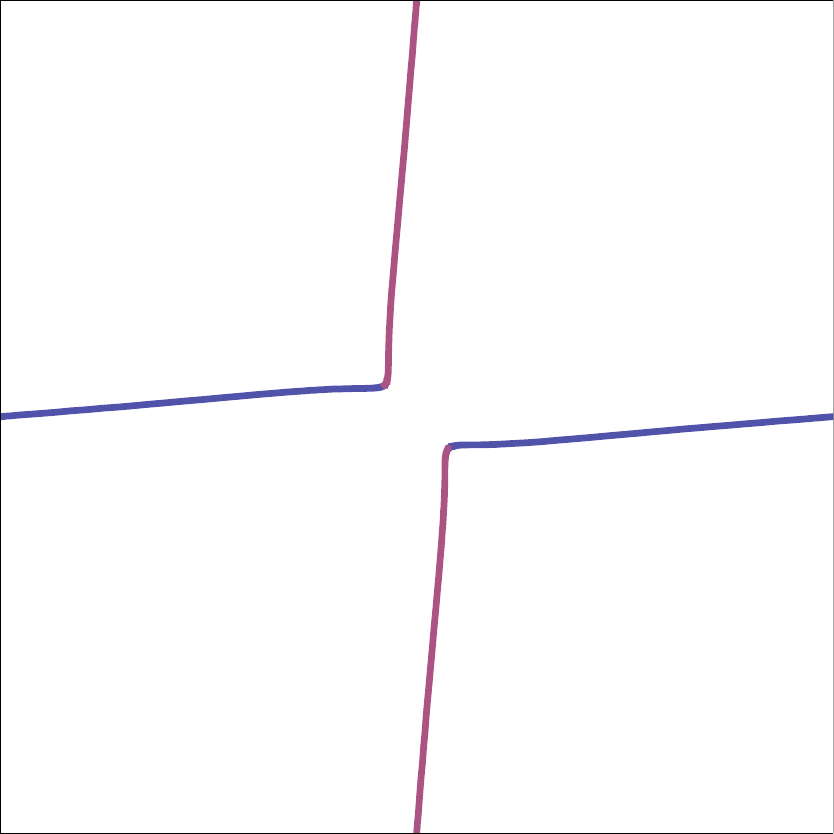}
\includegraphics[width=1.2in]{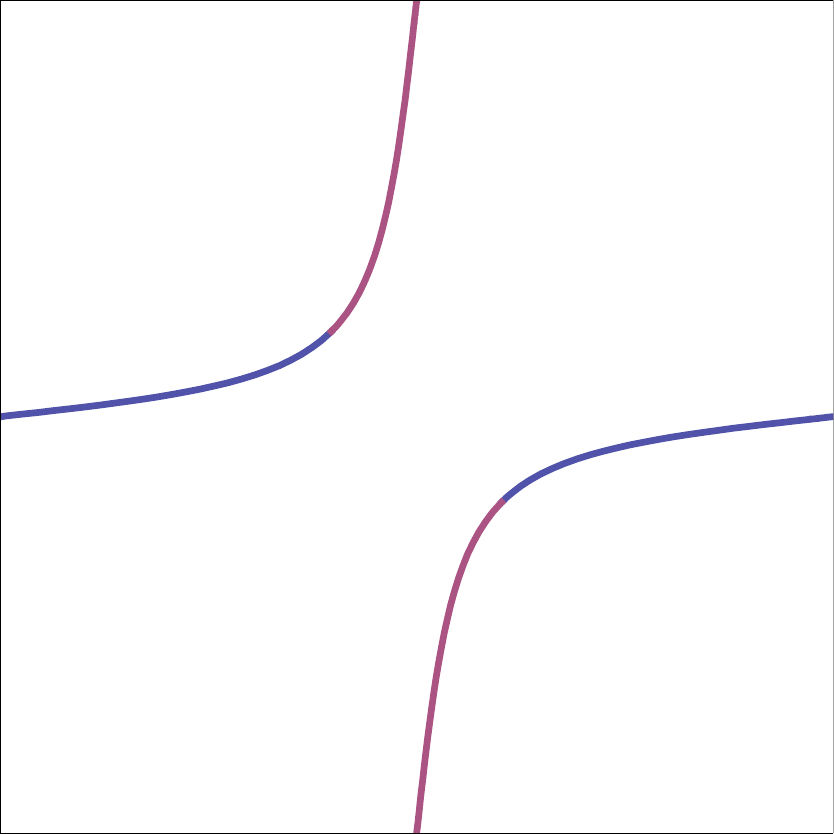}
\includegraphics[width=1.2in]{strecon3.pdf} \\
\includegraphics[width=1.2in]{strecon2.pdf}
\includegraphics[width=1.2in]{strecon1.pdf}
\includegraphics[width=1.2in]{strecon0.pdf}
\caption{String reconnection from the worldsheet and spacetime perspectives.  The upper panels show the equal-time contours before, during, and just after the reconnection process.  The lower panels show the corresponding spacetime trajectories.  We have labeled the portion of the contours in red (blue) corresponding to the asymptotic first (second) strings, but of course during reconnection the distinction is irrelevant since it is now one large string. } \ \\
\end{center}
\end{figure}

\subsection{Gravitational Radiation}
Now let us compute the radiation emitted from such a reconnection process.  The simplest and most physically relevant vertex operator is that for a graviton,
\begin{equation}
V_{\rm grav} =  \epsilon_{\mu \nu} \partial X_L^\mu {\bar \partial} X^\nu_R e^{-ik \cdot X_L/2 - ik \cdot X_R/2} 
\end{equation}
where we have factored the left- and right-moving components.  Our task now is to evaluate the amplitude
\begin{equation}
\label{amp}
\mathcal A(k) = \frac{1}{2 \pi \alpha'}  \int d^2 z \ V_{\rm grav}[X_{\rm cl}(z,{\bar z})] 
\end{equation}
given the trajectory (\ref{xcl}).  As explained in more detail in section 3.1, this problem is tantamount to determining the cosmic string's energy-momentum tensor
\begin{equation}
\label{tmunu}
 T^{\mu \nu}(k) = \frac{1}{2 \pi \alpha'} \int d^2 z \ \partial X^\mu_L {\bar \partial} X^\nu_R e^{-ik \cdot X_L/2 - ik \cdot X_R/2} . 
 \end{equation}
Damour and Vilenkin investigated this quantity for classical cosmic strings \cite{dandv}, finding that it is largest when both the left- and right-moving components of the exponential have either a saddle point or discontinuity at the same location.  The signal strength can then be conveniently ranked according to the phase behavior:
\begin{enumerate}
\item {\bf Cusp}: double saddle-point.
\item {\bf Kink}: one saddle-point, one discontinuity.
\item {\bf Reconnection}: double discontinuity.
\end{enumerate}
Since the contribution of radiation emitted from reconnection would appear to be a \emph{sub}-sub-leading correction, it is not surprising that henceforth no effort has been made to explicitly calculate this.  

The reason we believe this is now a worthwhile endeavor is because in
superstring theory the probability of reconnection $(\ref{p0})$ can be
greatly suppressed, $P \sim 0.01$.  As discussed in more detail in the
next section, this then produces a larger number of strings $\sim
1/P$, the number of reconnection attempts now scaling as $\sim 1/P^2$, and
the number of successful reconnections scaling as $ \sim 1/P$.  Thus, as
is the case for bursts from cusps, the lower
the probability of reconnection, the larger
the gravitational wave signal.

Let us make a naive attempt to evaluate the amplitude (\ref{tmunu}).  In examining $X_{\rm cl}$ in (\ref{xcl}), there do not appear to be any kink-like discontinuities, but there do appear to be cusps as indicated by the presence of extrema.  Focusing now on just the holomorphic part of $X_{\rm cl}$, the cusp condition is
\[0 = k \cdot \partial X_L(z)  = k \cdot p_{1L} \left( \frac{ 1}{z-\frac{1}{2}} +  \frac{ 1}{2-z} \right) +  k \cdot p_{2L} \left( \frac{1}{z+\frac{1}{2}} - \frac{ 1}{2+z} \right). \]
Rewriting this as a quadratic equation, one sees there are two solutions,
\[  z_\pm = \frac{5(1+r_L) \pm \sqrt{(9+r_L)(1+9r_L)}}{4(r_L-1)}, \hspace{0.5in} r_L = \frac{k \cdot p_{2L}}{k \cdot p_{1L}}. \]
One solution will be $z_- \sim 0$ and the other $z_+ \sim \infty$.  In the large-winding limit $L \rightarrow \infty$, away from these extrema the phase will oscillate wildly and so we attempt to utilize a saddle-point approximation,
\[ T^{\mu \nu} (k) \sim \left[ \frac{\partial X^\mu_L(z_-) e^{-ik \cdot X_L(z_-)/2} }{\sqrt{k \cdot \partial^2 X_L(z_-)}} + \frac{\partial X^\mu_L(z_+) e^{-ik \cdot X_L(z_+)/2} }{\sqrt{k \cdot \partial^2 X_L(z_+)}} \right] \times (L \rightarrow R, z \rightarrow {\bar z}) \hspace{0.3in} {\rm (naive)} . \]
This could not be possibly be correct, as this $T \sim L/k$ spectrum indicative of simultaneous cusps does not even remotely reproduce the $T \sim 1/k^2$ spectrum due to simultaneous kinks, which we must obtain in the classical limit.  Where did we go wrong?  The answer lies in the fact that our saddle-point evaluation was done at the edge of moduli space $z_+ \sim \infty$, where such an approximation breaks down \cite{Gross:1987ar}.

To evaluate the amplitude correctly, we follow Kawai-Lewellen-Tye \cite{Kawai:1985xq} and write the exponential part of the amplitude explicitly:
\[ \mathcal A(k) \sim \int d^2 z \left( \frac{z-\frac{1}{2}}{1-\frac{z}{2}} \right)^{\frac{\alpha'}{2} k \cdot p_{1L}} \left( \frac{z + \frac{1}{2}}{1+\frac{z}{2}} \right) ^{\frac{\alpha'}{2} k \cdot p_{2L}} \left( \frac{{\bar z}-\frac{1}{2}}{1-\frac{{\bar z}}{2}} \right)^{\frac{\alpha'}{2} k \cdot p_{1R}} \left( \frac{{\bar z} + \frac{1}{2}}{1+\frac{{\bar z}}{2}} \right) ^{\frac{\alpha'}{2} k \cdot p_{2R}} . \]
Denoting $z = x + iy$, the integrand is an analytic function of $y$ with eight branch points: $\pm i \left(x \pm \frac{1}{2} \right), \pm i \left( x \pm 2 \right)$.  The contour over $y$ stretches from $-\infty$ to $\infty$ along the real axis, so we can then rotate it counter-clockwise to now run from $-\infty$ to $\infty$ along the imaginary axis.  If we then define the \emph{independent} variables
\[ z_L \equiv x + iy, \hspace{0.5in} z_R \equiv x - i y \]
we can rewrite the integral in terms of these variables, paying special attention to the phases, 
\begin{eqnarray*} && \mathcal A(k) \sim \frac{1}{2} i \int_{-\infty}^\infty dz_L \int_{-\infty} ^\infty dz_R \left| \frac{z_L-\frac{1}{2}}{2-z_L} \right|^{ \frac{\alpha'}{2} k \cdot p_{1L}} \left| \frac{z_L + \frac{1}{2}}{2+z_L} \right| ^{\frac{\alpha'}{2} k \cdot p_{2L}} \left| \frac{z_R-\frac{1}{2}}{2-z_R} \right|^{\frac{\alpha'}{2} k \cdot p_{1R}} \left| \frac{z_R + \frac{1}{2}}{2+z_R} \right|^{ \frac{\alpha'}{2} k \cdot p_{2R}} \\
 & \times& e^{i \frac{\alpha'}{2} \{ \left[ \arg (z_L- \frac{1}{2})  - \arg(2-z_L) \right] k \cdot p_{1L} + \left[ \arg (z_L + \frac{1}{2})  - \arg(2+z_L) \right]  k \cdot p_{2L} - \left[ \arg (z_R- \frac{1}{2})  - \arg(2-z_R) \right] k \cdot p_{1R} - \left[ \arg (z_R + \frac{1}{2})  - \arg(2+z_R) \right] k \cdot p_{2R}  \} }.
\end{eqnarray*}
\begin{figure}
\begin{center}
\includegraphics[width=2.6in]{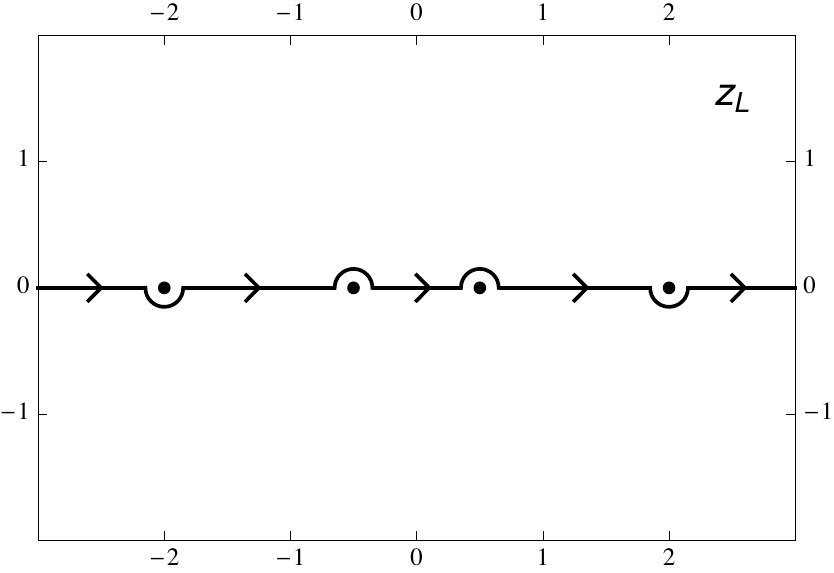}
\includegraphics[width=2.6in]{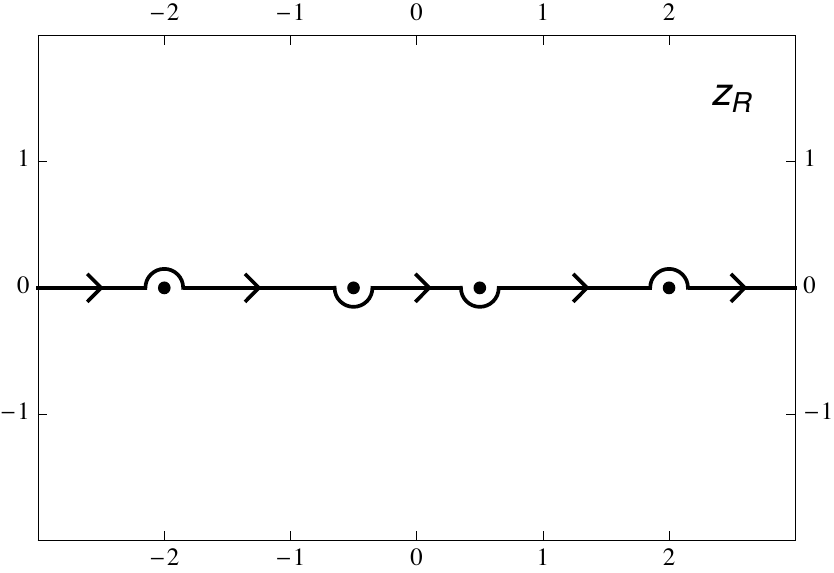}
\includegraphics[width=2.6in]{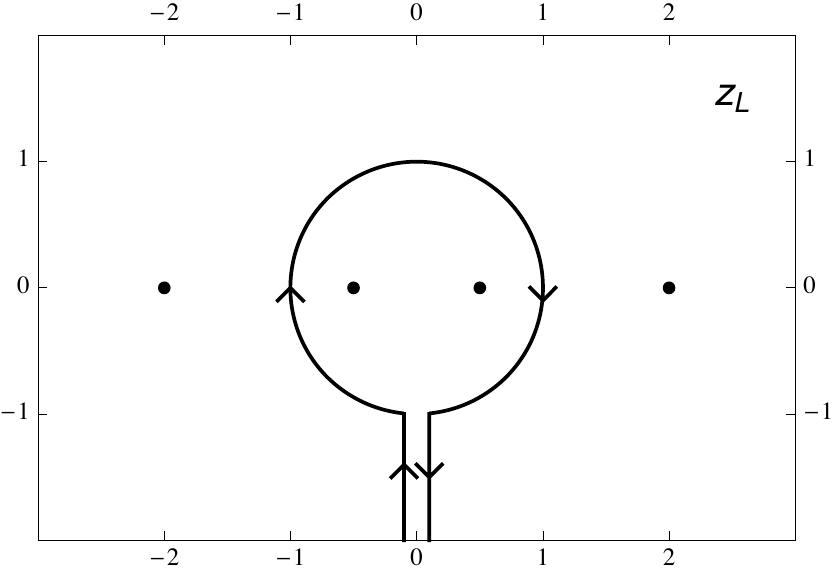}
\includegraphics[width=2.6in]{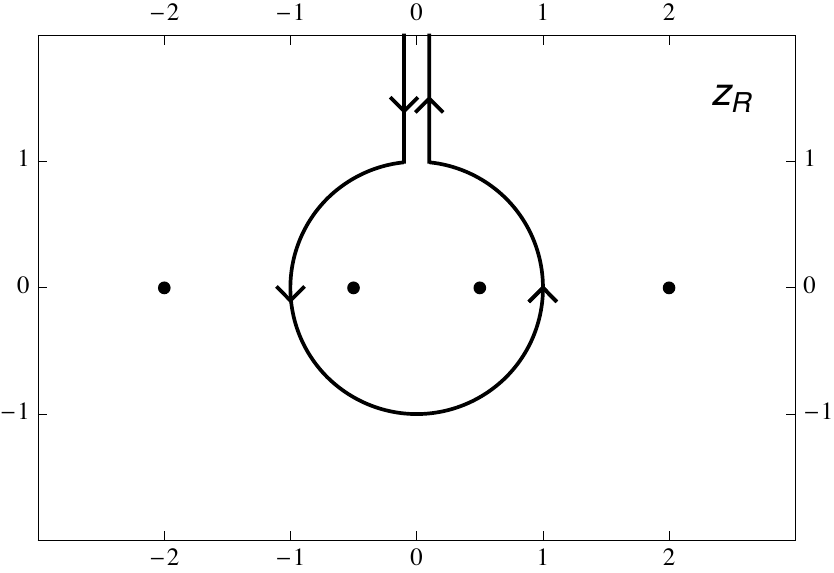}
\caption{The contours for $z_L, z_R$ initially lie on the real axis (above) and are chosen to circumnavigate the poles so as to produce the correct phases in the amplitude.  These can then be deformed into something more convenient for explicit evaluation, such as the complex unit circle with negligible contribution from infinity (below).} \ \\
\end{center}
\end{figure}
\begin{figure}
\begin{center}
\includegraphics[width=3in]{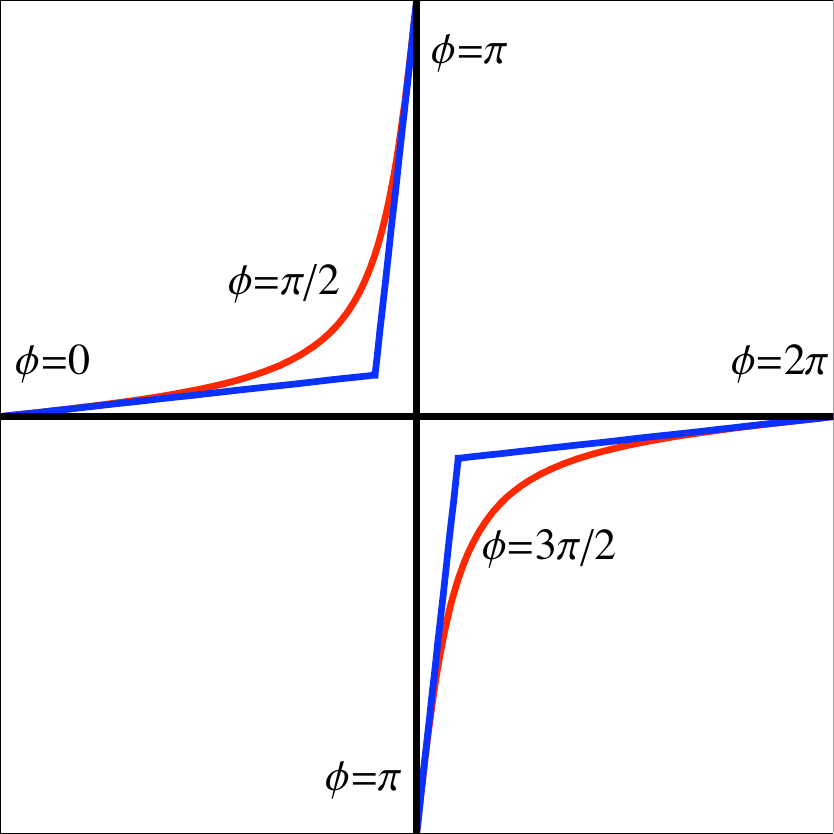}
\caption{The reconnection geometry for classical cosmic strings (black) versus cosmic superstrings, both exactly (red) and in the large-$L$ approximation (blue).} \ \\
\end{center}
\end{figure}
An examination of these phases produces the following summary:
\begin{center}
\begin{tabular}{|c|c|}
\hline
Region of $z_L$ & phase \\
\hline
$z_L < -2$ & $\frac{\alpha'}{2} \pi k \cdot p_{1L} $  \\
$-2 < z_L < - \frac{1}{2}$ & $\frac{\alpha'}{2} \pi \left( k \cdot p_{1L} + k \cdot p_{2L} \right) $ \\
$-\frac{1}{2} < z_L < \frac{1}{2}$ & $\frac{\alpha'}{2} \pi k \cdot p_{1L} $  \\
$\frac{1}{2} < z_L < 2$ & 0 \\
$2 < z_L$ & $ - \frac{\alpha'}{2} \pi k \cdot p_{1L} $ \\
\hline
\end{tabular}
\hspace{0.1in}
\begin{tabular}{|c|c|}
\hline
Region of $z_R$ & phase \\
\hline
$z_R < -2$ & $-\frac{\alpha'}{2} \pi k \cdot p_{1R} $  \\
$-2 < z_R < - \frac{1}{2}$ & $- \frac{\alpha'}{2} \pi \left( k \cdot p_{1R} + k \cdot p_{2R} \right)$ \\
$-\frac{1}{2} < z_R < \frac{1}{2}$ & $-\frac{\alpha'}{2} \pi k \cdot p_{1R} $  \\
$\frac{1}{2} < z_R < 2$ & 0 \\
$2 < z_R$ & $ \frac{\alpha'}{2} \pi k \cdot p_{1R} $ \\
\hline
\end{tabular}
\end{center}
These phases can be produced by choosing the contours shown in Figure 3.  Once these choices have been made, of course, we are free to deform them into something more convenient for analysis.  We choose to deform the contour into the complex unit circle, which we recall corresponds to zero time as shown in Figure 2.  Note that the trajectory at this moment resembles a smoothed-out form of the classical reconnection picture.  The idea of transforming an amplitude at a given point on the string over all time into over the entire string at a single moment in time is not without precedent in superstring theory \cite{Hashimoto:1996bf}.

Now that we have completely factored the left- and right moving components to (\ref{tmunu}), we are only concerned with the evaluation of the integral over the angle $\phi \equiv \arg z_L$,
\begin{equation}
\label{il}
I_L^\mu(k) = \int_{-\pi} ^\pi d\phi \ \partial_\phi X_L^\mu e^{-ik \cdot X_L(e^{i \phi})/2} .
\end{equation}
Defining the scale-invariant quantities ${\hat X}_L \equiv X_L/L, \psi_L \equiv L \phi_L$, we can expand (\ref{il}) in powers of $L$ near the points $z_L = \pm 1$ at which the trajectory is very nearly linear:
\begin{eqnarray}
\nonumber
I_L^\mu(k) &=&  \int_{-L \pi /2} ^{L \pi /2} d \psi \ \left[ \partial_\phi {\hat X}_L^\mu(1) + \frac{\psi}{L} \partial^2_\phi {\hat X}^\mu (1) + \cdots \right] e^{-ik \cdot \left[ \psi \partial_\phi {\hat X}_L(1) + \frac{\psi^2}{2L} \partial_\phi^2 {\hat X}_L(1) + \cdots \right] /2}  \\
\nonumber
&+& \int_{L \pi /2} ^{-L \pi /2} d \psi \ \left[ \partial_\phi {\hat X}_L^\mu(-1) + \frac{\psi}{L} \partial^2_\phi {\hat X}^\mu(-1) + \cdots \right] e^{-ik \cdot \left[ \psi \partial_\phi {\hat X}_L(-1) + \frac{\psi^2}{2L} \partial_\phi^2 {\hat X}_L(-1) + \cdots \right] /2} \\
\label{il2}
&\approx&  2i \left[  \frac{ \partial_\phi {\hat X}_L^\mu(1)}{k \cdot { \partial_\phi {\hat X}_L(1)}} -  \frac{ \partial_\phi {\hat X}_L^\mu(-1)}{k \cdot { \partial_\phi {\hat X}_L(-1)}} \right] \\
\nonumber
&=& 2i \left[ \frac{ {\hat p}_{1L}^\mu }{ k \cdot {\hat p}_{1L} - \frac{9}{80} \left[  k \cdot {\hat p}_{2L} +  \frac{(k \cdot {\hat p}_{1L})^2}{ k \cdot {\hat p}_{2L}} \right]}  - \frac{ {\hat p}_{2L}^\mu }{ k \cdot {\hat p}_{2L} - \frac{9}{80} \left[  k \cdot {\hat p}_{1L} +  \frac{(k \cdot {\hat p}_{2L})^2}{ k \cdot {\hat p}_{1L}} \right]}  \right] + \mathcal O \left( \frac{1}{k^2 L} \right). 
\end{eqnarray}
Thus in accordance with the classical expectation, we have obtained a spectrum which is scale-invariant in the large-winding limit, and which has virtually identical dependence upon the relative angle and velocity of the strings.  In the last line we have written it to most resemble the classical answer plus corrections; the terms proportional to $\frac{9}{80}$ are the corrections from the string path integral.  Of course the procedure is then repeated for the right-moving side.

We must also average over all geometric possibilities for which reconnection takes place using the probability measure
\begin{equation}
\label{pmeasure}
\int dP = \int_0^1 dv^2 \int _{-1} ^1 d (\cos \theta) \int _0 ^{2 \pi} d \phi \ P(v,\theta)
\end{equation}
modulated by some string distribution $n(v,\theta)$ and then finally to sum over polarizations.  While (\ref{pmeasure}) itself easily evaluates to $\int dP = \frac{5 \pi^3 g^2_s V_{\rm min}}{16 V_{\perp}} $, averaging (\ref{il2}) must be done numerically.

\subsection {Other Cosmic Superstring and Radiation Types}
Though we have considered massless gravitational modes, the same procedure will work for any string mode by using the appropriate vertex operator in (\ref{amp}).  However, analysis of the decay modes for highly-excited strings suggests that radiation into massive states is suppressed and that only massless quanta are excited \cite{Chialva:2003hg}.

In addition to the fundamental cosmic superstrings considered here, there are also D- and $(p,q)$-strings produced in brane inflation which will have significantly larger tension are therefore might be capable of emitting higher-energy bursts \cite{Schwarz:1995dk}  \cite{Jackson:2006qc}.  Though the method of calculation for reconnection interactions is very different \cite{Jackson:2004zg} \cite{Jackson:2007hn} \cite{Hanany:2005bc} \cite{Hashimoto:2005hi}, ideally their waveform could be estimated by adapting the F-string result (\ref{il2}) and appropriately replacing the inverse tension as $\alpha' \rightarrow \alpha'/ \sqrt{p^2+q^2/g_s^2}$ in the BPS limit, depending on the specific brane inflation model.  These $(p,q)$ strings also form more sophisticated networks \cite{pqnetworks} and thus would have distinctive burst patterns \cite{Brandenberger:2008ni}.

\subsection {Classical Cosmic String Reconnection}

For verification and comparison we now compute the string trajectory of two classical Nambu-Goto
strings during reconnection.  This is done using the geometry in Figure 1, taking the point of reconnection to be $\sigma = 0$ and the gauge conditions for the spatial coordinates ${\bf X}(\tau,\sigma)$ to be
\[ {\bf X} ^{\prime 2} +{ \dot {\bf X}}^2 = 1, \]
whereas the time coordinate is defined to be $X^0 \equiv 1$.  Though there exist analytic and numerical studies of the reconnection process \cite{vilshel} \cite{Achucarro:2006es}  \cite{Bettencourt:1994kc}, here we take the simplistic choice of instantaneous reconnection.   At time $\tau = 0$ we can write the spatial derivative and velocity for the first string as
\[ 
{{\bf X} ^\prime }_1 (0,\sigma) =
\left\{
\begin{array}{cc}
 \hat x & \sigma < 0 \\
\gamma^{-1} (\cos \psi  \hat x  + \sin \psi \hat y) & \sigma > 0
\end{array}
\right. ,
 \]
\[
\label{v1} 
\dot {\bf X}_1 (0,\sigma) =
\left\{
\begin{array}{cc}
 0 & \sigma < 0 \\
 \beta  \hat z & \sigma > 0
\end{array}
\right. ,
\]
and for string 2 as
\[
{{\bf X} ^\prime }_2 (0,\sigma) =
\left\{
\begin{array}{cc}
\gamma^{-1} (\cos \psi  \hat x  + \sin \psi \hat y) & \sigma < 0 \\
 \hat x & \sigma > 0
\end{array}
\right. ,
\]
\[
\dot {\bf X}_2  (0,\sigma) =
\left\{
\begin{array}{cc}
\beta  \hat z  & \sigma < 0 \\
 0 & \sigma > 0
\end{array}
\right. .
\]

Now defining $\sigma_\pm = \tau \pm \sigma$, we can always write ${\bf X}(\tau, \sigma) = {\bf X}_+(\sigma_+) + {\bf X}_-(\sigma_-)$ and then these tangent vectors combine into
\[
{\partial_\pm} {\bf X}_1 (\sigma_\pm) =
\left\{
\begin{array}{cc}
\pm {\hat x} & \pm \sigma_\pm < 0 \\
\beta {\hat z} \pm \gamma^{-1} (\cos \psi  \hat x  + \sin \psi \hat y) &  \pm \sigma_\pm > 0
\end{array}
\right. , \]
\begin{equation}
\label{dx2}
{\partial_\pm} {\bf X}_2 (\sigma_\pm) =
\left\{
\begin{array}{cc}
\beta {\hat z} \pm \gamma^{-1} (\cos \psi  \hat x  + \sin \psi \hat y)  & \pm \sigma_\pm < 0 \\
\pm {\hat x}& \pm \sigma_\pm > 0 .
\end{array}
\right. 
\end{equation}

These are then utilized in the evaluation of the integral (\ref{il}), which is easily done near a discontinuity:
\begin{eqnarray*}
I^\mu_\pm(k) &=& \int_{-\infty} ^\infty d \sigma_\pm \ \partial_\pm X^\mu e^{-ik \cdot X_\pm} \\
&\approx& 2i \left[ \frac{\partial X^\mu_\pm(0^+)}{k \cdot \partial X_\pm(0^+)} -  \frac{\partial X^\mu_\pm(0^-)}{k \cdot \partial X_\pm(0^-)} \right] .
\end{eqnarray*}

To compare these to the superstring results, recall that in (\ref{pl}) we have defined the left/right momentum vectors $p_{1L,R}, p_{2L,R}$, which can then be normalized as (substituting $+ \rightarrow L$, $- \rightarrow R$),
\[ {\hat p}^\mu_\pm \equiv \frac{p^\mu \pm \frac{ L^\mu}{2 \pi \alpha'}}{p^0} . \]
The classical result (\ref{dx2}) can then be directly compared to the superstring result (\ref{il2}) in terms of the geometry at the moment of reconnection, as shown in Figure 4.

\section {Detectability of bursts and stochastic background}

In a network of cosmic superstrings reconnections can occur in two
situations, (1) when two long strings meet, and (2) when loops are
formed.  The rate at which reconnections take place in a network of
superstrings, and the gravitational wave bursts that result, can be
readily estimated.
\subsection {Cosmic Superstring Detectability}
If we assume the string density to be inversely proportional to the
reconnection probability, at cosmic time $t$ the density of long
strings in the scaling regime is
\cite{Kibble:1976sj,vilshel,cstrings,Jackson:2004zg}
\[
\rho \sim \frac{\mu}{Pt^2},
\]
where $\mu$ is the mass per unit length in strings and $P$ is the
reconnection probability. The total energy in long strings is $\sim \rho t^3
\sim \mu t/P$ and the total length is $\sim t/P$, so the
total number of long strings in a Hubble volume at any time is $\sim
P^{-1}$ \cite{Sakellariadou:2004wq}.  Strings move relativistically so that in a time $t$ they travel a
distance $\sim t$ and each string attempts to reconnect with every
other string in a Hubble volume.  Although the number of reconnection
attempts in a time $t$ is $\sim P^{-2}$ the number of successful
reconnections is $\sim P^{-1}$. So the number of long string reconnections
that take place per unit space-time volume is
\begin{equation}
n_{R,S} \sim \frac{1}{Pt^4}.
\label{e:nR}
\end{equation}

Reconnections also occur when loops are formed. The scaling network
converts a length of string $\sim t/P$ into loops in a Hubble time
$\sim t$ per Hubble volume ($\sim t^3$).  If loops have a size
$\alpha t$ at formation, the number of loops formed in a time $t$ is $\sim (\alpha
P)^{-1}$, so the number of reconnections per unit space-time volume is
\begin{equation}
n_{R,L} \sim \frac{1}{\alpha Pt^4}.
\label{e:nL}
\end{equation}
If $\alpha$ is comparable to the Hubble length then the loop
contribution Eq.~(\ref{e:nL}) to the burst rate is comparable to the
string contribution Eq.~(\ref{e:nR}), but if the size of loops $\alpha$
is determined by gravitational back-reaction then the loop
contribution dominates. The total number of reconnections per unit
spacetime volume is thus,
\begin{equation}
n_R=n_{R,S}+n_{R,L} \sim \frac{1}{Pt^4}+\frac{1}{\alpha Pt^4} 
\sim \frac{1}{\alpha Pt^4}
\label{e:nL2}
\end{equation}
if $\alpha < 1$.  

The gravitational wave strain produced at the
reconnection event is of order
\[
h \sim \frac{G\mu}{rf^{2}},
\]
so that for a reconnection occurring at a redshift $z$,
\[
h(f,z) \sim \frac{H_0 G\mu}{\varphi_r(z)(1+z)f^2}.
\]
The amplitude of the waveform
is suppressed by an additional factor of $(1+z)$ due to the redshifting of
the frequency $f$ (see~\cite{dandv}).
The Hubble parameter today is $H_0=73$ km$/$s$/$Mpc \cite{hubble}), and we have
written the proper
distance as $r(z)=H_0^{-1} \varphi_r(z)$. 

Unlike cusps, these bursts do not point in any particular direction.
However, we do not observe bursts at arbitrarily low frequencies,
rather, only down to the inverse of the scale at which the strings curve ($\sim
\alpha t$) and the $f^{-2}$ behaviour of the waveform is no longer
valid.  Using Eq.~(\ref{e:nL2}) we can write the number of bursts
originating from a volume at redshift $z$ as
\[
\frac{dR}{dV(z)} \sim H_0^{4} \frac{(1+z)^{-1}}
{\alpha P\varphi_t^4(z)} \Theta((1+z)f\alpha H_0^{-1} \varphi_t(z) - 1)
\]
where the factor of $(1+z)^{-1}$ comes from the relation between the
observed burst rate and the cosmic time, and the $\Theta$-function cuts
off events that we would not observe below frequency $f$. We also take the volume
element at a redshift $z$ to be $dV(z)=H_0^{-3}\varphi_V(z)dz$ and so
the rate of events from redshifts between $z$ and $z+dz$ is
\begin{equation}
\frac{dR}{dz} \sim H_0 \frac{(1+z)^{-1}}{\alpha
  P\varphi_t^4(z)}\varphi_V(z)
\Theta((1+z)f\alpha H_0^{-1} \varphi_t(z) - 1).
\label{e:dRdz}
\end{equation}

To compute the observable rate of bursts we use the similar
methods to compute sensitivity estimates as used in~\cite{Siemens:2006vk}. 
We write the strain produced by the reconnection cusp as
\[
h(f,z) = A(z)f^{-2},
\]
with 
\begin{equation}
A(z) \sim \frac{H_0 G\mu}{(1+z)\varphi_r(z)}.
\label{e:amplitude1}
\end{equation}
Using the noise curve Eq.~(39) of~\cite{Siemens:2006vk} we find a
minimum detectable amplitude 
$A_{\rm min}=10^{-19}$~s for Initial LIGO and
$A_{\rm min}=10^{-20}$~s for Advanced LIGO.  Using
Eq.~(\ref{e:amplitude1}) we can find what redshifts these minimum
amplitudes correspond to for a given value of $G\mu$ ($z_{\rm max}$, say) and integrate
Eq.~(\ref{e:dRdz}) to find the rate
\[
R \sim \int_{0}^{z_{\rm max}} dz H_0 \frac{(1+z)^{-1}}{\alpha
  P\varphi_t^4(z)}
\varphi_V(z)\Theta((1+z)f\alpha H_0^{-1} \varphi_t(z) - 1).
\]

The cosmological functions that enter these expressions,
$\varphi_t(z)$, $\varphi_r(z)$, and $\varphi_V(z)$ depend on
cosmological parameters and need to be computed numerically as
described in Appendix A of~\cite{Siemens:2006vk}, where a vanilla
$\Lambda$-CDM model was used~\cite{hubble}. However, one can obtain
fairly good analytic approximations for the cosmological functions
using variations of the expressions introduced in~\cite{dandv}. In
particular the following expressions,
\[
\varphi_t=(1+z)^{-3/2} (1+z/z_{\rm eq})^{-1/2},
\]
with $z_{\rm eq}=5440$, 
\[
\varphi_r=z (1+z/3.5)^{-1},
\]
and
\[
\varphi_V=12z^2 (1+z/1.6)^{-13/2} (1+z/z_{\rm eq})^{-1/2}
\]
agree to better than 20\% with numerically computed functions at all
relevant redshifts. We will use these functions to compute the burst
rates and stochastic backgrounds.

\begin{figure}
\begin{center}
\includegraphics[width=6in]{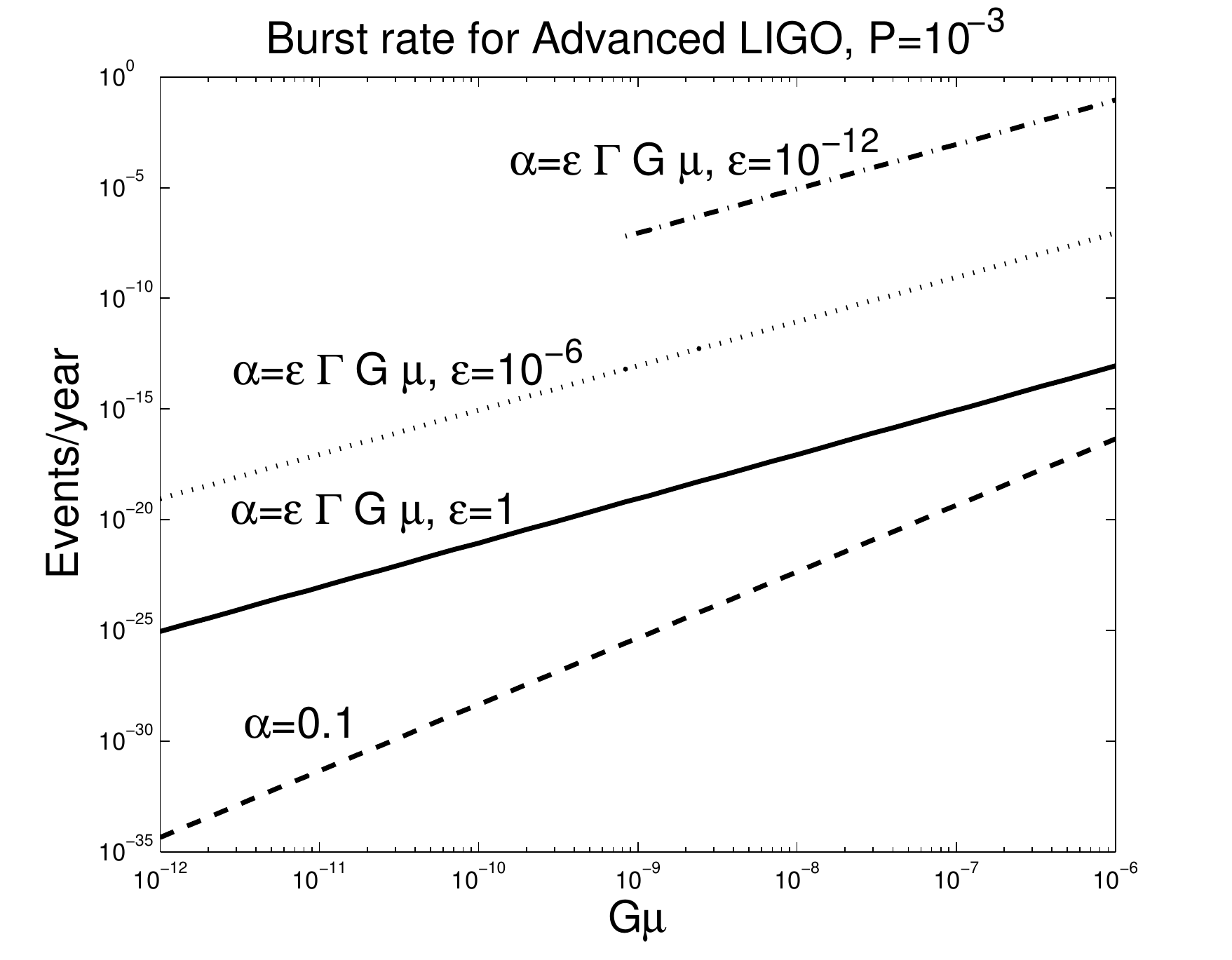}
\caption{Plot of the number of burst events per year for Advanced LIGO
  ($A_{\rm min}=10^{-20}$~s, $f=75$~Hz) as a function of
  string tension for a reconnection probability $P=10^{-3}$. The
  solid, dotted and dashed-dotted lines show the number of 
  events for loop sizes given by gravitational 
  back-reaction~\cite{Siemens:2001dx,Siemens:2002dj,Dubath:2007mf,Polchinski:2007rg,Polchinski:2006ee}
  $\alpha=\varepsilon \Gamma G \mu$ with $\varepsilon=1$,
  $\varepsilon=10^{-6}$, and $\varepsilon=10^{-12}$ respectively. We have taken
  $\Gamma=50$ everywhere.
  The dashed line shows the burst rate for loop sizes determined by
  the large scale dynamics of the network. We have taken $\alpha=0.1$
  as suggested by recent numerical
  simulations~\cite{Olum:2006ix}. Unfortunately, none of these models result in a
  signal detectable by LIGO.}
\label{fig:bursts} 
\end{center}
\end{figure}

Figure~\ref{fig:bursts} shows the number of burst events per year as a
function of string tension for for Advanced LIGO as a function of
string tension for a reconnection probability $P=10^{-3}$. We have
taken the minimum detectable amplitude to be $A_{\rm min}=10^{-20}$~s,
set the frequency to $f=75$~Hz, and taken $\Gamma=50$ everywhere. The
solid, dotted and dashed-dotted lines show the number of events for
loop sizes given by gravitational
back-reaction~\cite{Siemens:2001dx,Siemens:2002dj,Dubath:2007mf,Polchinski:2007rg,Polchinski:2006ee}
$\alpha=\varepsilon \Gamma G \mu$ with $\varepsilon=1$,
$\varepsilon=10^{-6}$, and $\varepsilon=10^{-12}$ respectively. The dashed
line shows the burst rate for large long lived loops ($\alpha = 0.1$) as
suggested by recent numerical simulations~\cite{Olum:2006ix}.
Unfortunately, none of these models result in a signal detectable by
LIGO.

To compute the stochastic background of gravitational waves generated
by all the reconnections occurring in the network we can use the
results in \cite{Siemens:2006yp} for cusps. The spectrum of
gravitational waves is given by
\begin{equation}
\Omega_{\rm gw}(f) = \frac{4 \pi^2}{3H^2_0}f^3
\int_{z_{\rm min}}^{\infty} dz \, h^2(f,z)  \frac {dR}{dz}
\label{e:Omega(f)2}
\end{equation}
where $z_{\rm min}$ excludes large bursts that occur infrequently and
contaminate the estimate of the strain~\cite{dandv,Siemens:2006yp},
and $h(f,z)$ is the strain produced at frequency $f$ by a burst from a
reconnection at redshift $z$.

\begin{figure}
\begin{center}
\includegraphics[width=6in]{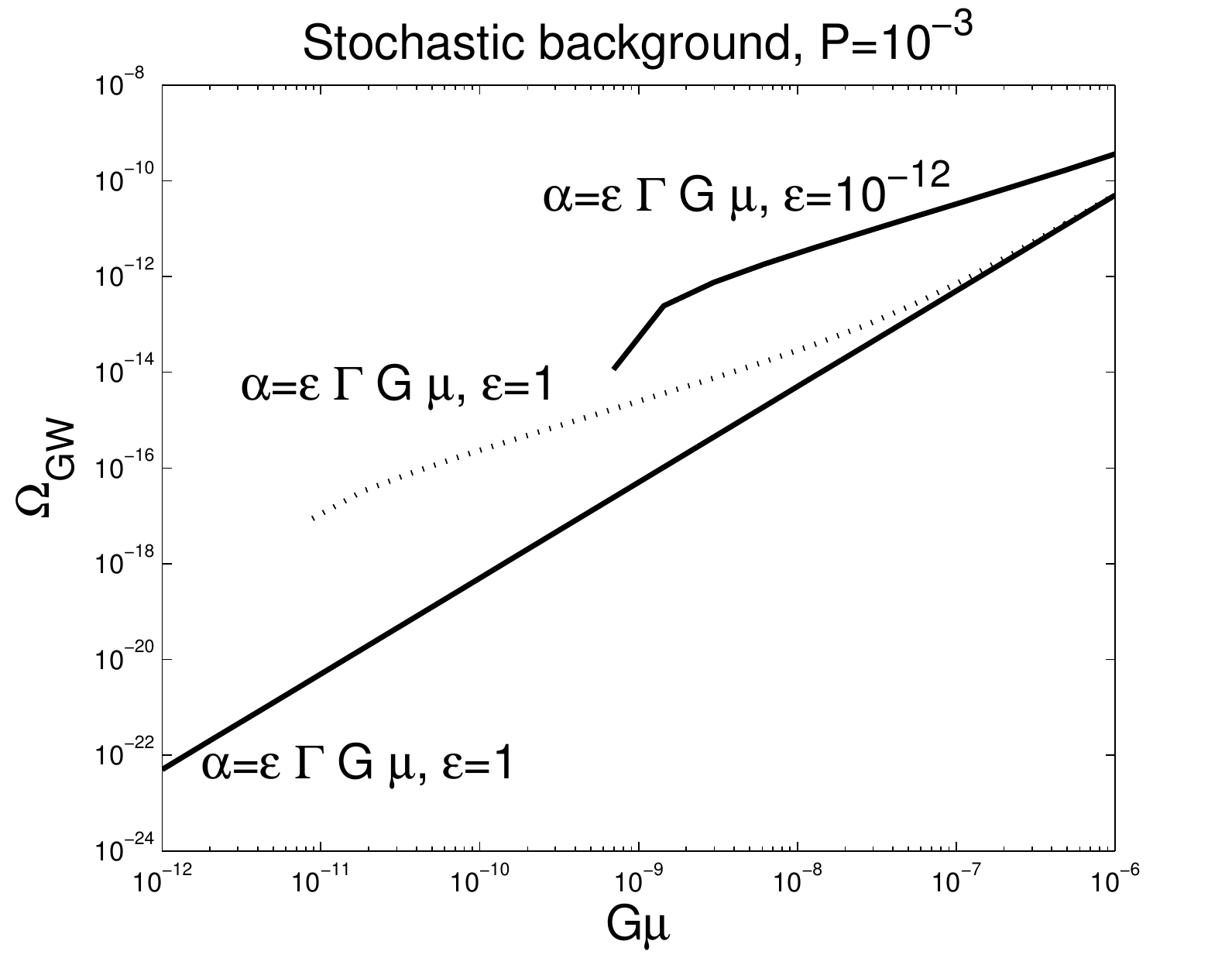}
\caption{Plot of $\Omega_{\rm GW}$
as a function of the string
tension for $p=10^{-3}$ and $\alpha=\varepsilon \Gamma G
\mu$.  The solid lines show the stochastic background generated at
$75$~Hz, corresponding to LIGO frequencies, the dotted line shows the
spectrum at $10^{-8}$~Hz, the band at which pulsar timing experiments
are most sensitive. Again, unfortunately, none of these models result 
in a detectable signal. Although not shown here, large loops
($\alpha=0.1$) also do not result in a detectable signal.}
\label{fig:stoch} 
\end{center}
\end{figure}

Figure~\ref{fig:stoch} shows three examples of the value of $\Omega_{\rm
  GW}$ at the frequencies of various experiments as a function of the
string tension for $P=10^{-3}$.  The solid lines show the stochastic
background generated at $75$~Hz, corresponding to LIGO frequencies,
for two values of the loop size.  The dotted line shows the spectrum
at $10^{-8}$~Hz, the band at which pulsar timing experiments are most
sensitive. Again, unfortunately, none of these models result in a
detectable signal. Although not shown here, large loops ($\alpha=0.1$)
do not result in a detectable signal.

\section {Discussion and Conclusion}

We have calculated the gravitational waveform due to cosmic
superstring reconnection, finding that bursts do not occur frequently
enough to be seen with upcoming gravitational wave detectors, nor is
the stochastic background strong enough to be detected.  This
conclusion remains true even for very auspicious values of currently
unknown parameters such as reconnection probability and loop sizes.
Thus the most likely source of cosmic string gravitational bursts
remains that resulting from cusps or kinks.

This is doubly disappointing since one very attractive feature of the
spectrum is that the gravitational radiation is polarized along the
directions of $(v,\theta)$ at reconnection and so there must be a
consistency relation with the probability distribution $P(v,\theta)$
encoded in the kinks on horizon-sized strings.  LISA, however, may be
sensitive enough to detect bursts from reconnections and a detailed
calculation including confusion noise from the network is underway.

Despite this disappointing result from gravitational bursts, the
techniques we have developed here can immediately be generalized to
other (non-gravitational) particles, which could be observed in
complementary experiments such as gamma-ray bursts or other
astrophysical phenomena.  Since the suppression is only power-law and
not exponential in energy, there remains the possibility of producing
very high energy particles.

\section{Acknowledgments}

We would like to thank A.~Ach\'ucarro, J.~J. Blanco-Pillado, T. Damour, G. Dvali, M. Sakellariadou and A. Stebbins for useful discussions, and especially J.
Polchinski and A. Vilenkin for finding a critical mistake in the first draft of this article.  MGJ
is supported by the DOE at Fermilab and would like to thank the
organizers of the Strings and Superstrings in Observational Cosmology
Workshop at APC-Paris.

\appendix
\section{Appendix: Gravitational radiation formalism}
In the weak field limit the metric is assumed to be that of a perturbed
Minkowski space
\[
g_{\mu\nu}=\eta_{\mu\nu}+h_{\mu\nu},
\]
with $|h_{\mu\nu}| \ll 1$.  In the harmonic gauge, defined by
\[
\partial_{\mu} h^{\mu}_{\nu}-{1\over 2}\partial_{\nu} h^{\mu}_{\mu}=0,
\]the linearised Einstein equations read
\begin{equation}
\label{lineq}
\partial_\lambda \partial^\lambda h_{\mu \nu}
=-16\pi G (T_{\mu \nu}-{1\over 2}\eta_{\mu\nu}T^{\sigma}_{\sigma})
\end{equation}
where $T_{\mu\nu}$ is the stress-energy tensor of the source. The retarded potential solution to Eq.~(\ref{lineq}) is given by
\[
h_{\mu\nu}({\bf x},t)=4G \int {d^3{\bf x}' \over |{\bf x}-{\bf x}'|}
(T_{\mu \nu}({\bf x}',t-|{\bf x}-{\bf x}'|)
-{1\over 2}\eta_{\mu\nu}T^{\sigma}_{\sigma}({\bf x}',t-|{\bf x}-{\bf x}'|)).
\]
For sources localised in space and time we can construct the
stress-energy tensor from the Fourier integral
\[
T_{\mu\nu}({\bf x},t)=\int_{-\infty} ^{\infty} d\omega \ T_{\mu\nu}({\bf x},\omega) e^{-i\omega t},
\]
with the inverse transformation given by
\begin{equation}
\label{Fourierinverse}
T_{\mu\nu}({\bf x},\omega)={1\over 2 \pi }\int_{-\infty} ^{\infty} dt \ T_{\mu\nu}({\bf x},t) e^{i\omega t}.
\end{equation}
The retarded potential solution is
\begin{equation}
\label{retardedFourierint}
h_{\mu\nu}({\bf x},t)=4G \int_{-\infty} ^{\infty} d\omega
\int {d^3{\bf x}' \over |{\bf x}-{\bf x}'|}
(T_{\mu \nu}({\bf x}',\omega)
-{1\over 2}\eta_{\mu\nu}T^{\sigma}_{\sigma}({\bf x}',\omega))
 e^{-i\omega (t-|{\bf x}-{\bf x}'|)}.
\end{equation}

We replace $|{\bf x}-{\bf x}'|$ with $r=|{\bf x}|$ in the
denominator of the integrand in Eq.~(\ref{retardedFourierint}) and
in the exponent approximate $|{\bf x}-{\bf x}'|\approx
r-{\hat \Omega} \cdot {\bf x}'$. This yields
\[
h_{\mu\nu}({\bf x},t) \approx {4G \over r} \int_{-\infty} ^{\infty}d\omega \
 e^{-i\omega (t-r)}
\int d^3{\bf x}' \
(T_{\mu \nu}({\bf x}',\omega)
-{1\over 2}\eta_{\mu\nu}T^{\sigma}_{\sigma}({\bf x}',\omega))
 e^{-i\omega_n {\hat \Omega} \cdot {\bf x}'}.
\]
If we now substitute for $T_{\mu \nu}({\bf x},\omega)$
using Eq.(\ref{Fourierinverse}) we get
\begin{eqnarray*}
\label{retardedFourierapproxint2}
h_{\mu\nu}({\bf x},t) &\approx& {4G \over r} \int_{-\infty} ^{\infty} d\omega \
 e^{-i\omega (t-r)}
\nonumber
\\
\nonumber
&\times&
{1\over 2\pi}\int_{-\infty} ^{\infty} dt\int d^3{\bf x}' \
(T_{\mu \nu}({\bf x}',t)
-{1\over 2}\eta_{\mu\nu}T^{\sigma}_{\sigma}({\bf x}',t))
 e^{i\omega (t-{\hat \Omega} \cdot {\bf x}')}.
\end{eqnarray*}
We can thus write
the solution as the integral over the plane waves
\[
h_{\mu\nu} ({\bf x},t)=\int_{-\infty} ^{\infty} d\omega \
\epsilon_{\mu\nu}({\bf x},\omega) e^{-ik \cdot x},
\]
where
\[
 \epsilon_{\mu\nu} ({\bf x},\omega)={4G \over r} (
T_{\mu \nu}(k)-{1\over 2} \eta_{\mu \nu} T^\lambda _\lambda(k) )
\]
is the polarisation tensor, and
\[
T^{\mu \nu }(k)=\frac{1}{2\pi}
\int_{-\infty} ^{\infty} dt \int d^{3}{\bf x} \ T^{\mu \nu }(x)e^{ik \cdot x}
\]
is the Fourier transform of the stress-energy tensor.


\end{document}